\documentclass{epl}
\usepackage{graphicx}
\usepackage{epsfig}
\usepackage{amsmath}
\usepackage{amssymb}

\title{Fluctuating lattice Boltzmann} 

\author{$^1$R. Adhikari, $^2$K. Stratford, $^1$M.E. Cates and $^3$A.J. Wagner}
\institute{$^1$School of Physics and $^2$EPCC, University of Edinburgh, JCMB
Kings Buildings, Mayfield Road, Edinburgh EH9 3JZ GB; $^3$Dept of Physics,
North Dakota State University, Fargo ND58105, USA}

\begin{document}
\maketitle

\begin{abstract} The lattice Boltzmann algorithm efficiently simulates the
Navier Stokes equation of isothermal fluid flow, but ignores thermal
fluctuations of the fluid, important in mesoscopic flows. We show how to adapt
the algorithm to include noise, satisfying a fluctuation-dissipation theorem
(FDT) directly at lattice level: this gives correct fluctuations for mass and
momentum densities, and for stresses, at all wavevectors $k$. Unlike previous
work, which recovers FDT only as $k\to 0$, our algorithm offers full
statistical mechanical consistency in mesoscale simulations of, e.g.,
fluctuating colloidal hydrodynamics.  \end{abstract}

The lattice Boltzmann equation (LBE) is a widely used lattice formulation of
fluid mechanics \cite{succi}. It offers a faithful discretization of the Navier
Stokes equation of isothermal, incompressible fluid flow, and is very well
adapted to parallel computation \cite{parallel}. While used for large-scale
fluid dynamics simulations such as flows around aircraft \cite{aircraft}, the
LBE approach is particularly adapted to simulating mesoscopic problems
\cite{discreteproc}. These include, e.g., porous medium flows and flows of
complex and multicomponent fluids with microstructure
\cite{warren,kendon,coveney,ladd}. The latter can be modelled using various
extensions of the basic algorithm for a single component fluid as considered
here \cite{swift,coveneyalg,ladd}. 

However, the Navier Stokes equation, and with it the LBE, ignores thermal
fluctuations. While these may safely be ignored in macroscopic fluid-dynamical
flows, at mesoscopic length scales they form an essential part of the physics
\cite{landau-fm}.  This applies even in linear problems such as the Brownian
motion of a colloidal particle suspended in a simple fluid: if that fluid is
simulated using the LBE, no Brownian motion occurs \cite{ladd}.  Fluctuations
are also central to nonlinear phenomena such as mode-coupling effects and
long-time tails \cite{mct}. By the same token, extensions of the LBE to fluid
mixtures \cite{swift} and amphiphilic solutions \cite{coveneyalg} cannot
address critical phenomena, where fluctuations dominate. 

In this letter we present a fluctuating LBE (FLBE). This offers a fully
consistent discretization of the equations of fluctuating nonlinear
hydrodynamics for an isothermal fluid, opening the way to more accurate and
efficient simulation of many of the mesoscale physics problems mentioned
previously, such as colloid hydrodynamics. Its generalization to multicomponent
fluids is conceptually straightforward; we pursue this elsewhere \cite{future}.
Our work also raises broader issues for numerical statistical mechanics: how
best to implement fluctuation-dissipation theorems (FDTs), derived in the
continuum with respect for appropriate conservation laws, in a system
discretized in space and time \cite{rivet}. We contend that accuracy and
efficiency are best combined if FDT is made to hold {\em directly} on the
discretized dynamics.

We start with the LBE for a single-component fluid, which can be viewed as a
discretization of the Boltzmann equation for the collisional dynamics of a
dilute gas \cite{benzi1}:
\begin{equation}\label{LBE}
f_i({\bf x}+{\bf c}_i,t+1)=f_i({\bf x},t)+{\mathcal L}_{ij}(f_j({\bf x},t)-f_j^{0}({\bf x},t))
\end{equation}
Here $f_i$ represents the local mass density of particles in a phase space cell
$({\bf x},{\bf c}_i)$, and is normalized so that $\sum_{i=1}^n f_i = \rho({\bf
x})$, the fluid mass density at ${\bf x}$. The spatial coordinates $\bf{x}$ are
discretized onto a unit lattice; a finite set of $n$ velocities ${\bf c}_i$ is
chosen so that in timestep $\Delta t = 1$, the resulting `streaming'
displacements $\Delta{\bf x}_i = {\bf c}_i\Delta t$ are lattice vectors. The
local momentum density and flux are $g_{\alpha}=\sum_i f_i c_{i\alpha}$ and
$\Pi_{\alpha\beta} =\sum_if_i c_{i\alpha} c_{i\beta}$; Greek indices indicate
Cartesian directions. The equilibrium distribution $f_i^{0}$ is conditioned by
the local values of $\rho$, $g_{\alpha}$ and $\Pi_{\alpha\beta}$.  The
collision operator $\mathcal{L}_{ij}$ linearly relaxes the local phase-cell
densities towards this equilibrium. For well chosen lattices and collision
operators, the LBE is known to recover the isothermal Navier-Stokes equation in
the continuum limit at low Mach number \cite{succi}. 

The $f_i$ in Eq.\ref{LBE} are {\em ensemble-averaged} local distribution
functions.  Although these depend on temperature (indeed, $f_i^{0}$ for a fluid
at rest is Boltzmann distributed \cite{mb-note}), they describe only the
\emph{mean} densities in each phase space cell. Accordingly, they neglect the
fluctuating stress $s_{\alpha\beta}$ in the equations of fluctuating nonlinear
hydrodynamics \cite{landau-sp1} for an isothermal fluid:
\begin{subequations}
\label{FNH}
\begin{eqnarray}
\label{MASSCONSERVATION}
\partial_t \rho &+& \nabla_{\alpha}g_{\alpha}=0,\\
\label{MOMENTUMCONSERVATION}
\partial_t g_{\alpha}&+&\nabla_{\beta}\Pi_{\alpha\beta}=0,\\
\label{CONSTITUTIVE}
\Pi_{\alpha\beta}=g_{\alpha}v_{\beta}+p\delta_{\alpha\beta}&
+&\eta_{\alpha\beta\gamma\delta}\nabla_{\gamma}v_{\delta}+s_{\alpha\beta}.
\end{eqnarray}
\end{subequations}
In Eqs.\ref{FNH}, $v_{\alpha}=g_{\alpha}/\rho$ is the local fluid velocity, $p$
is the pressure in a quiescent fluid (given here by an ideal gas equation of
state $p=\rho c_s^2$, with $c_s$ the isothermal sound speed), and
$\eta_{\alpha\beta\gamma\delta}$ 
is a tensor of viscosities. The fluctuating stress 
$s_{\alpha\beta}$ is a zero-mean Gaussian
random variable whose variance, for a fluid at temperature $T$, is fixed by the
fluctuation-dissipation theorem (FDT) to be $\langle s_{\alpha\beta}({\bf
x},t)s_{\gamma\delta}({\bf
x'},t')\rangle=2k_BT\eta_{\alpha\beta\gamma\delta}\delta({\bf x-x'})\delta(t-t')$.

An important precursor to our work is that of Ladd \cite{ladd}. His method
consists of adding a stochastic piece to the microscopic stress tensor in
Eq.\ref{LBE}. Following through to the continuum limit, Eq.\ref{FNH} is
recovered.  However, this process ensures {\em only} that FDT holds in the
hydrodynamic limit $k\to 0$. In practice, no numerical algorithm
is ever used in this limit, which would require infinite computational
resources. Indeed, in the noiseless LBE, acceptable hydrodynamic behavior of
the fluid is maintained right up to $k\simeq 2$ \cite{behrend}; accordingly
accurate colloid hydrodynamics is achieved with colloids of rather small
radius, e.g., $R = 2.5$ \cite{ladd}. (Lengths are expressed in lattice units, so that $k=2$ has a wavelength of $\pi$).  But unless the
correct noise behavior is implemented over a similarly wide $k$ range,
breakdown of FDT at high $k$ is liable to infect the whole simulation, once
noise is added. We show below, for a simple benchmark problem of colloids in traps, that this can
indeed occur, resulting in errors of order 10\%. 
(These errors might be reduced by making $R$ several times larger at fixed volume fraction; but
this is not computationally efficient.)
Moreover, as found below, such errors show nontrivial dependence on
parameters such as the colloid volume fraction. Without exhaustive testing, therefore, one cannot know whether Ladd's algorithm is accurate or not, in any given region of parameter space. 
In contrast, our own algorithm appears to give good equilibration of colloidal degrees of freedom whenever the suspending fluid is accurately in equilibrium.

To make progress, we note that in addition to the hydrodynamic degrees of
freedom appearing in Eq.\ref{FNH}, the LBE necessarily involves the dynamics of
local, non-hydrodynamic modes often called `ghosts' \cite{vergassola,benzi2}.
These are needed in Eq.\ref{LBE} to maintain isotropic and Gallilean invariant
hydrodynamics \cite{benzi1,d'Humi'eres}. As shown below, the method of
Ref.\cite{ladd} effectively sets $T=0$ for the ghost mode noise.  There is then a risk that ghosts drain thermal
fluctuations away from the hydrodynamic modes, which therefore may never reach
equilibrium. 

To create our improved FLBE, we promote Eq.\ref{LBE} into a discrete Langevin
equation, where the $f_i$ are interpreted as instantaneous, fluctuating
densities in phase space:
\begin{equation} \label{FLBE}
f_i({\bf x}+{\bf c}_i,t+1)=f_i({\bf x},t)+{\mathcal L}_{ij}(f_i({\bf x},t)-f_i^{0}({\bf x},t))+\xi_i
\end{equation}
with noise terms $\xi_i({\bf x},t)$ that give fluctuations in the populations
in each phase space cell. To recover thermal equilibrium, the $\xi_i$ {\em
must} be linked, by an FDT, to all sources of dissipation in the collision
operator $\mathcal{L}_{ij}$.  The required FDT must allow that, at a given site
and timestep, the $\xi_i$ are correlated in such a way as to exactly conserve
$\rho$ and $ g_{\alpha}$. To recover continuum hydrodynamics, the collision
process necessarily {\em avoids} relaxing the conserved quantities, which
propagate only via the `streaming step' ${\bf x} \to {\bf x} + {\bf c}_i$ in
Eq.\ref{LBE} \cite{succi}. Below we diagonalize the collisional dissipation,
using established methods \cite{d'Humi'eres}. In contrast to previous work,
we determine the accompanying noise structure consistent
with statistical mechanical principles.

First, note that each unnormalised eigenvector $m_{i}^{a}$ of
$\mathcal{L}_{ij}$with eigenvalue ($-1/\tau_a$) is associated with a
corresponding local density $M^{a}({\bf x},t)$ via                  
\begin{equation}\label{DECOMP1}
M^{a}({\bf x},t)=\sum_{i=1}^n m^{a}_i f_i({\bf x},t), a=1\ldots n
\end{equation}
For example, the density $\rho({\bf x},t)$ equates to $M^{1}({\bf x},t)$ where
$m^{1}_i = 1\, \forall\, i$. For a
general LBE in $d$ dimensions containing $n$ velocities (a `D$d$Q$n$
model'), there are precisely $n$ eigenvectors, corresponding to the $n$ degrees
of freedom contained in the $f_i$ at a given site.  Each $M^a$ relaxes at
a rate given by the eigenvalue $-1/\tau_a$. Conserved hydrodynamic
variables (mass, momentum, but not stress) do not relax; we set the
corresponding eigenvalues to zero ($-1/\tau_a = 0$) without loss of
generality. Note that our method allows for a multi-relaxation time kernel but does not assume it;
ghost noise is important even with a single relaxation 
time.

A complete mode count then consists of the null eigenvector $m^{1}_i$
corresponding to the density $\rho$; the $d$ null eigenvectors $c_{i\alpha}$
corresponding to the $d$ components of the momentum $g_{\alpha}$;
$\frac{1}{2}d(d+1)$ eigenvectors
$Q_{i\alpha\beta}=c_{i\alpha}c_{i\beta}-c_s^2\delta_{\alpha\beta}$
corresponding to (independent components of) the deviatoric momentum flux
$S_{\alpha\beta}=\Pi_{\alpha\beta}- \nu k_BT\delta_{\alpha\beta}$, where
$\nu$ is the number density; and the remaining
$n-(1+d+\frac{1}{2}d(d+1))$ ghost mode eigenvectors. The latter are
model-dependent, but readily computed from $\mathcal{L}_{ij}$ in any given
implementation of the LBE \cite{succi}.  
Our $n$ eigenvectors form a complete, orthogonal basis in the finite velocity space
of the LBE. Thus 
\begin{equation}\label{DECOMP2}
f_i({\bf x},t)=\sum_a w_i m^{a}_i M^{a}({\bf x},t){N^a}
\end{equation}
with normalizers $N^a$ obeying
$N^a\sum_i w_i m^{a}_i m^{b}_i = \delta_{ab}$
and 
$w_i$ a set of known weights \cite{succi}.

A similar expansion can be applied to any function defined on the velocity space,
including the noise $\xi_i({\bf x},t)$. Let us first set
$\xi_i = \xi_i^H + \xi_i^G$ with $H$ the hydrodynamic subspace and $G$ is its
complement, the ghost subspace. Identifying the 
$H$ eigenvectors as above, we write
\begin{equation} 
{\xi_i^H}= \sum_{a\in H} w_i m^{a}_i \hat\xi^{a}({\bf x},t)N^a
= w_i\left(m^{1}_i{\hat\rho} + {c_{i\alpha}{\hat g_{\alpha}}\over c_s^2}
+ {Q_{i\alpha\beta}{\hat S_{\alpha\beta}}\over 2c_s^4}\right)
\label{EXPFULL}\end{equation}
where the noise terms $\hat\xi^{a}$ associated with $\rho, g_\alpha, S_{\alpha\beta}$ are
denoted by $\hat\rho$...; repeated Greek indices are summed on.

Within this framework, conservation laws have a very simple effect: mass and
momentum conservation demand $\hat\rho = 0$ and $\hat g_\alpha = 0$ in
Eq.\ref{EXPFULL}. We are left with 
\begin{equation}\label{EXP}
\xi_i=w_i{Q_{i\alpha\beta}{\hat S_{\alpha\beta}({\bf x},t)}\over 2c_s^4} + 
\sum_{g\in G} w_i m^{g}_i \hat\xi^{g}({\bf x},t)N^g
\end{equation}
Here the first term is $\xi_i^H$: it produces thermal fluctuations in the
stress tensor, and is the noise used by Ladd \cite{ladd}.  The remaining terms
are $\xi_i^G$: these maintain thermal equilibrium for the ghosts, and are new
to this work. Our approach differs crucially from earlier work on continuous and discrete stochastic versions of the Boltzmann equation \cite{bixon,dufty}; these attempted to derive the noise statistics from the collision kernel alone, without taking proper account of mass and momentum conservation.

It remains to determine the covariance matrix of the $n-d-1$ nonzero noises
$\langle \hat\xi^{a}\hat\xi^{b}\rangle$ that now appear in
Eqs.\ref{EXPFULL},\ref{EXP}. 
Projecting Eq.\ref{FLBE} onto the basis of eigenvectors,
taking deviations $\delta M^a$ around a spatially uniform state and Fourier
transforming in space we obtain
\begin{equation}\label{MOMENT-FLBE}
\delta M^{a}({\bf k},t+1) = \Gamma^{ab}({\bf
k})\left\{r^b\delta M^{b}({\bf k},t) + \hat\xi^{b}({\bf k},t)\right\} 
\end{equation}
where $\Gamma^{ab}({\bf k})=\sum_{i=1}^{n}w_i m_i^a m_i^b N^{b}\exp(-i{\bf
k}\cdot{\bf c}_i)$ and $r^b=(1-\tau^{-1}_{b})$. Eq.\ref{MOMENT-FLBE} represents
a set of coupled Langevin equations in discrete time. We now assume that all
correlations are {\bf k}-independent; this is justified below.  Next we square
both sides of the above equation, average over the noise, and invoke
stationarity of equal-time correlators to obtain the required FDT:
\begin{equation}
\label{FDT}
\left\langle{\hat\xi^{a}\hat\xi^{b}}\right\rangle=\frac{\tau_a+\tau_{b}-1}
{\tau_a\tau_{b}}\left\langle \delta M^{a} \delta M^{b}\right\rangle 
\end{equation}
The $\tau$-dependence is a standard consequence of the discrete
time dynamics in Eq.\ref{FLBE} \cite{succi,ladd}.  The $d+1$ null modes
corresponding to conserved quantities have no dissipation and
hence no noise.  All remaining modes, including ghosts,
have both thermal fluctuations and dissipation; for consistent
dynamics, they {\em must} have noise. Note from Eq.\ref{MOMENT-FLBE} that setting
$\tau_b=1$ for ghosts \cite{ladd}, does {\em not} decouple the hydrodynamic
modes from the ghost noise, except strictly at ${\bf k}={\bf 0}$.

To complete our calculation of the noise amplitudes, we now quantify the
equilibrium thermal fluctuations $\langle\delta M^{a}\delta M^{b}\rangle$.
This requires a thermodynamic model for our fluctuating fluid. In keeping with
the original thinking behind the LBE \cite{succi}, and also with its practical
application (low Mach number), we can choose for this model the thermodynamics
of an ideal gas. The fluctuation matrix $\langle \delta M^{a}\delta
M^{b}\rangle$ is then ${\bf k}$ independent, as promised above, and computable
from knowing that all equilibrium phase-cell occupancies in such a gas obey
Poisson statistics \cite{landau-sp1}.  The required matrix $\langle \delta
M^{a}\delta M^{b}\rangle$ then follows by a change of basis from $f_i$ to
$M^a$, using Eq.\ref{DECOMP1}.  In implementing our FLBE numerically, we
instead transform in the other direction via Eq.\ref{DECOMP2}. This gives from
Eq.\ref{FDT} a set of correlated phase-cell noises $\xi_i({\bf x},t)$ for use
in Eq.\ref{FLBE}. As shown above, the noises acting on the $n$ different
densities $f_i$ at site ${\bf x}$ are not independent: they derive from only
$n-d-1$ underlying independent noises.  And, as our eigen-analysis of the
collision operator makes clear, they arise from the collisional dissipation of
the ${1\over2}d(d+1)$ stress modes {\em and} the
$\left(n-(1+d+{1\over2}d(d+1)\right)$ ghost modes.  The computational overhead
of adding ghost noise is slight ($\sim$ 10\% on run time).

\begin{figure}
\begin{center}
\begin{tabular}{cc}
\includegraphics[width=4.1cm]{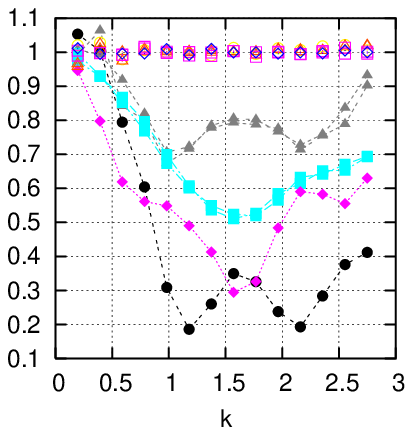}&
\includegraphics[width=4.1cm]{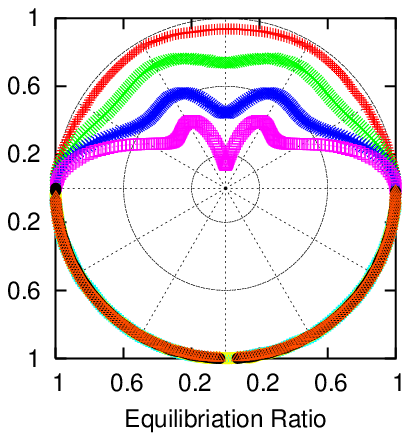}\\
\end{tabular}
\caption{\label{firstfig} Left panel: Equilibration ratios for the Fourier
modes of the density ($\circ$), momentum ($\triangle$), and diagonal
($\square$) and off-diagonal ($\lozenge$) stress components in a D3Q15 model
with $\tau = 0.75$ shown as a function of $k$ with ${\bf k}= k(1,1,1)$.  We set
$k_BT=1/3000$. Open symbols: the present work; closed symbols: setting
$\xi_i^G = 0$. (Compare Fig.3(d) of \cite{behrend} where the correct
hydrodynamics, without noise, is shown to hold up to $k\sim 2$.) Right panel:
$\langle |g_x({\bf k})|^2\rangle$ in a $D2Q9$ model with $\tau=1$, plotted as a
function of $\theta$ for various values of $k$, when ${\bf
k}=k(\cos\theta,\sin\theta)$. The radial coordinate is the equilibriation ratio
with unity as the solid circle. Lower half plane: the present work (note that
symbols overlap). Upper half plane: setting $\xi_i^G = 0$. Top to bottom at
upper right, $k = 0.62, 1.25, 1.88, 2.51$.}
\end{center}
\end{figure}

The consistency of our FLBE can be assessed by measuring numerically the
`equilibration ratio' (ER) for fluctuating hydrodynamic quantities. This is the
ratio of a measured variance to the one required by the Boltzmann distribution
\cite{mb-note} at the temperature $T$ chosen for the simulation. We set mass,
length, and time units so that $\rho=1$ on an unit lattice. In \emph{any} quiescent fluid 
the net thermal momentum $\tilde g$ in volume $\Delta V$ has variance
$\langle \tilde g^{2}\rangle = \Delta V \rho k_BT$ \cite{landau-fm}. For the 
ideal gas $k_BT = p/\nu = \rho c_s^2/\nu = c_s^2/\nu$, where $\nu$ fixes the phase-cell occupancies referred to above, and hence
all noise amplitudes. Setting $\Delta V = 1$, the on-site thermal velocity $v=\tilde g/\rho$ obeys $\langle v^2\rangle = k_BT$. Thus, we must have $k_BT \ll c_s^2 = 1/3$ (for
D$d$Q$n$ models) to satisfy the low Mach number requirement of the LBE. 

We now report results for $D2Q9$ and $D3Q15$ lattices, using a collision
operator with unit relaxation time for ghost modes and relaxation time $\tau$
for stress.  The viscosity tensor is that of an isotropic fluid, with shear
viscosity $\eta=c_s^2(\tau-{1\over2})$ and bulk viscosity $\zeta={2\over
d}\eta$. (This is a particular case of the multirelaxation time operator used
in \cite{ladd}.) Fig.\ref{firstfig} compares ERs for different hydrodynamic
modes on a $D3Q15$ lattice, and shows the dependence on direction in wavevector
space of the momentum fluctuations on a $D2Q9$ lattice.  We also show the
results found by setting $\xi_i^G = 0$ in Eq.\ref{EXP}, as per Ref.\cite{ladd}.
As anticipated above, the latter gives acceptable ERs only for $k\ll 1$, with
values of $0.2-0.7$ for $1<k<2$, although the noiseless hydrodynamics remains
accurate here \cite{behrend}. In contrast, our algorithm is accurate
throughout. 

\begin{figure}
\begin{center}
\includegraphics[width=5cm]{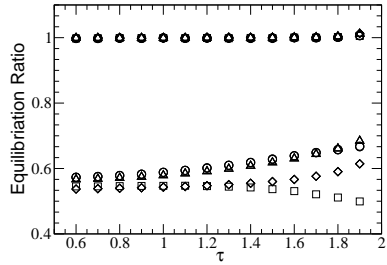}
\end{center} \caption{\label{secondfig} Equilibration ratios for on site
density ($\square$), momentum ($\circ$) and stress ($\bigtriangleup$ diagonal, $\lozenge$ off-diagonal) fluctuations as a function of the relaxation time
$\tau$ in a $D3Q15$ model.  We set $k_BT=1/3000$.  Upper datasets: the
present work (note that symbols overlap). lower
datasets: setting $\xi_i^G = 0$.}
\end{figure}

Figure \ref{secondfig} shows on-site ERs for various fluctuating quantities as
a function of $\tau$. These involve integration over all $k$, and could be
viewed as a `worst case' scenario for the local temperature seen by a small
colloidal particle. Our algorithm is accurate, even in this limit. We recommend choosing $k_BT \le 1/3000$ so that the fluid is accurately incompressible; we
then find equilibration at the one percent level or better.

To simulate colloids directly, we have combined our noise implementation with
the algorithm of \cite{laddBB} for the colloidal particles \cite{kevinfuture}.
We find results for velocity autocorrelators that offer a slight improvement
over the case with ghost noise omitted \cite{codef}; but for standard parameter
settings both methods give acceptably accurate values for the colloidal
diffusion constant $D = k_BT/(6\pi\eta R)$.
A more demanding test of equilibration is to simulate colloids in harmonic
traps, of the kind often encountered in experiments with optical
tweezers \cite{traps}. 
We placed $N$ particles, each in a separate confining
potential, at regular intervals on an $L^3$ lattice. The particles interact
hydrodynamically, but this does not affect the Boltzmann distribution
for their thermal displacements within (well-separated) traps.

\begin{figure}
\begin{center}
\includegraphics[height=0.25\textwidth]{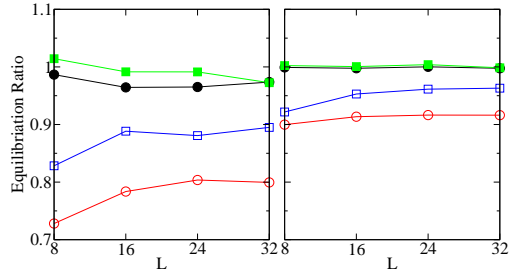}
\end{center} \caption{\label{thirdfig} Equilibration ratios for colloidal displacements in traps as a function of periodic box size $L$ for $\eta = 1/6$ (left) $\eta = 1/60$ (right) and two colloidal radii (circles $R = 1.5$,  squares $R=2.3$) in a $D3Q15$ model.  We set $k_BT=1/19200$. Upper datasets: the
present work; lower datasets, the same but setting $\xi_i^G = 0$. Error bars are of order the symbol size.
}
\end{figure}

A statistically decisive comparison can be made for $N = 1$, when a given particle is surrounded not by other colloids in independent traps but by its own periodic images, with which it interacts hydrodynamically. In figure \ref{thirdfig} we show ERs
for the thermal displacement of such a particle. 
Whereas our results are satisfactory, Ladd's algorithm shows worsening ERs as $L/R$ is decreased \cite{pcladd}. Any attempt to circumvent these large
($>10\%$) systematic errors arising from omission of ghost noise by defining an
`effective temperature' would merely move those errors into $D$ instead. FDT violations in the ghost sector seemingly present major obstacles to the
accurate simulation of colloids, particularly at modest interparticle 
spacings (modest $L/R$) \cite{laddBB} which are overcome by our algorithm \cite{footn}.

We have also performed tests with $N > 1$ (now fixing $L = 32$). The ER is found for each
particle separately by running for $10^6$ timesteps with $\eta=1/60$ and
$k_BT=1/3000$. (The equilibration time within a trap is of order $r_0^2/D \sim 10^4$.)
For $N=64$ particles of $R=2.3$ in traps where the correct rms thermal
displacement is $r_0 = \sqrt{3}$, we find ERs of $0.861\pm0.018$ for Ladd's
algorithm and $0.957\pm0.017$ for our own. For $R = 2.3$ and $r_0 = \sqrt{3}/2$ ($N=64$) the results are $0.928\pm0.007$ versus our $1.055\pm0.011$. For $R = 6.23$ and $r_0 = \sqrt{3}/2$ ($N=16$) the results are $0.962\pm0.011$ versus our $1.020\pm0.015$ \cite{error}. 
These findings with $N>1$ suggest broadly improved results with our algorithm. However a complete exploration of parameter space would be needed to fully resolve the differences of the two methods: without this, one cannot know in advance when it is safe to neglect ghost noise. This situation is made worse for dense colloids without traps; here the confining potential created on a colloid by its
neighbours might not be sampled correctly, but that cannot be checked by explicit calculation. 

In conclusion, by writing a discrete Langevin equation at lattice level and
carefully applying a fluctuation dissipation theorem, we have derived a
fluctuating lattice Boltzmann equation in which the ghost modes are fully
thermalized. This removes their fluctuation-draining
effects on the hydrodynamic sector at finite $k$. Our method gives improved
numerical equilibration of fluctuating quantities at all $k$, thus
resolving a potentially major obstacle to the use of lattice Boltzmann methods in the
simulation of thermal fluids, including colloid hydrodynamics.

Acknowledgements: We thank Ignacio Pagonabarraga for useful discussions. We thank Tony Ladd for many useful
discussions and for showing us his own results on traps \cite{pcladd}. Work funded by EPSRC GR/R67699 (RealityGrid) and GR/S10377. 


\begin{thebibliography}{}
\bibitem{kevinfuture} Our numerics follow Ladd's `bounce-back' methodology
\cite{ladd,laddBB}. See also \cite{codef}.
\bibitem{footn} For our recommended range of $k_BT \le 1/3000$, colloidal equilibration remains good to two or three percent. 
At much higher temperatures, ERs significantly larger than unity arise in the fluid and (for uncertain reasons) are amplified further in the colloid sector.
\bibitem{error} We quote standard errors of the mean but note that the statistical error might be larger in some cases due to long-lived hydrodynamic correlations;
we find these to be weak for our parameters.
\end{thebibliography}
\end{document}